\documentclass[12pt,english]{article}\usepackage[]{graphicx}\usepackage{xcolor}
\makeatletter
\def\maxwidth{ %
  \ifdim\Gin@nat@width>\linewidth
    \linewidth
  \else
    \Gin@nat@width
  \fi
}
\makeatother

\definecolor{fgcolor}{rgb}{0.345, 0.345, 0.345}

\usepackage{framed}
\makeatletter
 {\par\unskip\endMakeFramed%
 \at@end@of@kframe}
\makeatother

\definecolor{shadecolor}{rgb}{.97, .97, .97}
\definecolor{messagecolor}{rgb}{0, 0, 0}
\definecolor{warningcolor}{rgb}{1, 0, 1}
\definecolor{errorcolor}{rgb}{1, 0, 0}
\newenvironment{knitrout}{}{} 

\usepackage{alltt}

\usepackage[a4paper]{geometry}
\usepackage{algorithm}
\usepackage{graphicx}
\usepackage{alphalph}
\usepackage[utf8]{inputenc}
\usepackage[T1]{fontenc}
\usepackage{tikz}
\usepackage{float}
\usepackage[colorlinks = true,linkcolor = blue, citecolor = blue]{hyperref}
\usepackage{booktabs}
\usepackage{amssymb,amsmath}
\usepackage{natbib}

\def\aaa{a}
\def\aaag{\textbf{a}}
\def\R{\mathbb{R}}

\def\diag{{\rm diag}}
\def\E{{\rm E}}

\def\varepsilong{\boldsymbol{\varepsilon}} 
\def\pig{\boldsymbol{\pi}} 

\def\Db{{\bf D}}

\def\ab{{\bf a}}

\def\yb{{\bf y}}
\def\zb{{\bf z}}

\def\ub{{\bf u}}
\def\xb{{\bf x}}
\def\vb{{\bf v}}

\def\Ab{{\bf A}}
\def\Bb{{\bf B}}
\def\Cb{{\bf C}}
\def\Mb{{\bf M}}

\def\Ib{{\bf I}}
\def\Xb{{\bf X}}

\def\Db{{\bf D}}

\def\Sb{{\bf S}}

\def\Wb{{\bf W}}
\def\0b{{\bf 0}}
\def\1b{{\bf 1}}

\newtheorem{exmpl}{Example}[section]

\newcommand{\myalphafoot}
{
\renewcommand{\thefootnote}{\alph{footnote}}
}

\title{Spatial Spread Sampling\\Using Weakly Associated Vectors}
\myalphafoot
\author{\myalphafoot Rapha\"el Jauslin\footnotemark[1]~ and Yves Till\'e\footnotemark[1]}
\date{}
\footnotetext[1]{Institute of statistics, University of Neuchatel, Av. de Bellevaux 51, 2000 Neuchatel, Switzerland\\ (E-mail: raphael.jauslin@unine.ch)}

\makeatletter
\DeclareRobustCommand{\change}{%
  \@bsphack
  \leavevmode
  \color{red}%
  \@esphack
}

\DeclareRobustCommand{\stopchange}{%
  \@bsphack
  \normalcolor
  \@esphack
}
\makeatother

\IfFileExists{upquote.sty}{\usepackage{upquote}}{}
\begin{document}

\maketitle


\begin{abstract}
Geographical data are generally autocorrelated. In this case, it is preferable to select spread units.
In this paper, we propose a new method for selecting well-spread samples from a finite spatial population with equal or unequal inclusion probabilities. The proposed method is based on the definition of a spatial structure by using a stratification matrix.
Our method exactly satisfies given inclusion probabilities and provides samples that are very well-spread.
A set of simulations shows that our method outperforms other existing methods such as the Generalized Random Tessellation Stratified (GRTS) or the Local Pivotal Method (LPM). Analysis of the variance on a real dataset shows that our method is more accurate than these two. Furthermore, a variance estimator is proposed.
\end{abstract}


\newpage
\section{Introduction}

Data from natural resource surveys are often spatially autocorrelated, meaning that two close measurements are similar. In general, to estimate a total of a target variable, selecting the units spatially best spread collect more information and provides better estimation. An important problem of spatial sampling is thus to spread at best the sampled units in space. A well-spread sample is called spatially balanced. \citet{graf:lund:13} and \citet{gra:sch:14} give the formal definition of a representative sample and discuss the theoretical justification of taking a well-spread sample with unequal probabilities.
\citet{marker2009sampling} and \citet{hank:mohr:newm:2020} present some example of studies where the population considered is in an environmental context such as lakes, wetlands, rangelands, and forests. \citet{vall:till:2015} discuss forest ecosystem evolution using a well-spread spatial sampling design. \citet[][Chapter~8]{till:2020}, \citet{till:wilh:2017}, \citet{Bene:Pier:Posti:2017} and  \citet{Wang20121} give a review of the main spatial sampling methods. \citet{que:49a} and \citet{bellhouse1977some} showed that systematic sampling is the optimal design for autocorrelated data.

Generalized Random Tessellation Stratified (GRTS) sampling is a spatial sampling method proposed by \citet{stevens1999spatially,Stev:Olse:vari:2003,Stev:Olse:spat:2004}.
They use a mapping by means of a quadrant-recursive function to map a finite subset of a multi-dimensional space into the real line. A one-dimension systematic sampling is then applied, possibly with unequal probabilities \citep[see also][]{theo:07,brown2015spatially,spsurvey}. \citet{rob:mcd:pri:brown:18} have proposed a similar method called Halton iterative partitioning (HIP). It uses structural properties of the Halton sequence to draw a well-spread sample.
\citet{dick:till:2016} have simply used the Traveling Salesman Problem (TSP) in order to map the population points in one dimension. Systematic sampling is then applied. \citet{gra:11} has proposed spatially correlated Poisson sampling (SCPS). This method uses weights to create strong negative correlations between the inclusion probabilities of nearby units. \citet{gra:lun:sch:12} proposed the Local Pivotal Method (LPM). It is a particular case of the splitting methods proposed by \citet{dev:til:98}. It consists of randomly choosing between two nearby units at each step and produces an automatic repulsion in the selection of the neighbour units. \citet{gra:til:13} have generalized the LPM to obtain spread samples that are also balanced on totals of auxiliary variables. All these methods are implemented in the BalancedSampling {R} package \citep{graf2019}.

\citet{Stev:Olse:spat:2004} have proposed to compute the Vorono{\"\i} polygons around the sampled units, after which they sum the inclusion probabilities of the population units belonging to each Vorono{\"\i} polygon. The variance of these sums, called ``spatial balance'', is an indicator of the quality of spreading. \citet{til:dic:esp:giu:18} have modified the index proposed by \citet{moran1950notes} so that it can be interpreted as a coefficient of correlation between the units and their neighbourhood. The index provides another measure of the quality of spreading.

\cite{dig:10} defined preferential sampling as a sample selection where the sampling method is not independent of the spatial process, and where unequal inclusion probabilities cannot be explained by auxiliary variables. It is important to emphasize that, in this manuscript, the inclusion probabilities are supposed to be established in advance. The sample selection is a random realization of the sampling model and is independent of all of the variables.

In this paper, we propose a new spatial sampling method. We start with the vector of inclusion probabilities. Like in the cube method \citep{dev:til:04a,til:06} inclusion probabilities are randomly modified at each step. It can be seen as random walk that from the vector of inclusion probabilities ends up with a sample. By choosing well the modification direction at each step the sample selected is very well-spread.

The paper is organized as follows. Section~\ref{sec:setup} gives the notation and a basic setup of the problem as well as the insight that a well-spread sample results in an Horvitz-Thompson estimator with a smaller variance. In Section~\ref{sec:wave}, we introduce the new method that we propose and the process of sample selection. In Section~\ref{sec:index}, we describe the indices that enable to evaluate the quality of the spreading: the spatial balance index and the measure based on Moran's $I$ index. In Section~\ref{sec:variance}, we present a variance estimator for our method.
In Section~\ref{sec:artificial}, we give simulation results of the algorithm on artificial spatial configurations while Section~\ref{sec:realdata} is dedicated to simulations on real data. We used the geo-referenced ``Meuse'' dataset available in the R package ``sp'' of \citet{pebe:biva:2005} with inclusion probabilities proportional to the ``cadmium'' variable. Simulations show that the proposed method surpasses, LPM, GRTS and SCPS for the quality of the spreading, and the estimation accuracy.


\section{Notation}\label{sec:setup}
\subsection{Basic setup}

Consider a finite population $U$ of size $N$ whose units can be defined by labels $k\in\{1,2,\dots,N\}$. Let $\mathcal{S} = \{s | s\subset U\}$ be the power set of $U$. These units are geo-referenced in a space that can have more than two dimensions. A sampling design is defined by a probability distribution $p(.)$ on $\mathcal{S}$ such that

$$
p(s) \geq 0 \text{ for all } s\in \mathcal{S} \text{ and }\sum_{ s\in \mathcal{S}}p(s) = 1.
$$

A random sample $S$ is a random vector that maps elements of $\mathcal{S}$ to an $N$ vector of 0 or 1 such that $\textrm{P}(S = s) =
p(s)$. Define $a_k(S)$, for $k = 1,\dots,N$:

$$
\aaa_k =
\left\{\begin{array}{lll} 1 & \text{ if } k\in S\\ 0 & \text{ otherwise} . \end{array} \right.
$$

Then a sample can be denoted by means of a vector notation:
$ \aaag^\top = (\aaa_1,$ $\aaa_2,$ $\dots,$ $\aaa_N).$ For each unit of the population, the inclusion probability $0\leq\pi_k\leq 1$ is defined as the probability that unit $k$ is selected into sample $S$:
\begin{equation*}\label{eq:pik}
 \pi_k = \textrm{P}(k \in S) = \textrm{E}(\aaa_k) =  \sum_{s\in S | k \in s} p(s), \text{ for all } k\in U.
\end{equation*}

Let $\pig^\top=(\pi_1,\dots,\pi_N)$ be the vector of inclusion probabilities. Then, $\textrm{E}({\aaag})=\pig.$ In many applications, inclusion probabilities are such that samples have a fixed size $n$. Let the set of all samples that have fixed size equal to $n$ be defined by
 \begin{equation*}\label{eq:sn} \mathcal{S}_n = \left\{ \aaag\in \{0,1\}^N ~~\bigg|~~ \sum_{k
= 1}^N \aaa_k = n \right\} .
 \end{equation*}

The sample is generally selected with the aim of estimating some population parameters. Let $y_k$ denote a real number associated with unit $k\in U$, usually called the variable of interest. For example, the total
$$
Y=\sum_{k\in U} y_k
$$
can be estimated by using the classical Horvitz-Thompson estimator of the total defined by
\begin{equation}\label{eq:HT}
\widehat{Y}_{HT} = \sum_{k\in U} \frac{y_k a_k}{\pi_k}.
\end{equation}

Usually, some auxiliary information $\xb_k^\top = (x_{k1},x_{k2},\dots,x_{kq}) \in\mathbb{R}^q$  regarding the population units is available.
In the particular case of spatial sampling, a set of spatial coordinates  $\zb_k^\top = (z_{k1},z_{k2},\dots,z_{kp}) \in \mathbb{R}^p$ is supposed to be available, where $p$ is the dimension of the considered space. A sampling design is said to be balanced on the auxiliary variables $x_k$ if and only if it satisfies the balancing equations
\begin{equation*}\label{eq:balance}
  \widehat{\Xb} = \sum_{k\in S} \frac{\xb_k}{\pi_k} = \sum_{k\in U} \xb_k = \Xb.
\end{equation*}


\subsection{Well-spread sample}\label{sec:spatjust}
A sample is well spread ``if the number of selected units is close to what is expected on average in any part of the space'' \citep{graf:lund:13}. We give in this section an insight that selecting a well-spread sample minimizes the variance of the Horvitz-Thompson estimator. Suppose we are in the general linear superpopulation model:

$$ \yb_k = \xb_k^\top\beta + \varepsilon_k, ~ \text{ for all } k \in U,$$
where $\xb_k$ is a column vector of values taken by $q$ auxiliary variables on unit $k$, $\beta\in\R^q$ are $q$ regression coefficients and $\varepsilon_k$ is a random variable that satisfies $\E_M(\varepsilon_k) = 0$ and $\text{var}_M(\varepsilon_k) = \sigma^2(\xb_k) = \sigma_k^2$, with $\sigma^2(\cdot)$ a Lipschitz continuous function. Note that $E_M(\cdot)$ and $\text{var}_M(\cdot)$ are the expectation and the variance under the model. Let also

$$ \text{cov}_M(\varepsilon_k,\varepsilon_\ell) = \sigma_k\sigma_\ell \rho_{k\ell},~~ \text{ with } k\neq \ell \in U,$$
where $\rho_{k\ell}$ is a function that decreases when the distance between two units increase. This notation shows that two close units are autocorrelated. \citet{gra:til:13} showed that

\begin{equation}\label{eq:spatjust} \E_p\E_M( \widehat{Y}_{HT} - Y)^2 = \E_p\left[\left(\sum_{k\in S}\frac{\xb_k}{\pi_k} -\sum_{k\in U}\xb_k \right)^\top\beta\right]^2 + \sum_{k\in U}\sum_{\ell\in U} \sigma_k\sigma_\ell \rho_{k\ell} \frac{\pi_{k\ell} - \pi_k\pi_\ell}{\pi_k\pi_\ell},
\end{equation}
where $\E_p$ is the expectation of the design and $\pi_{k\ell} = \E_p(a_k a_\ell)$ is the joint inclusion probabilities. From equation \eqref{eq:spatjust} we could see that the first term of the right hand side is minimized if the sample is balanced on the auxiliary variables $\Xb$. The second term is minimized if $\pi_{k\ell}$ is small whenever $\rho_{k\ell}$ is large. Meaning that choosing a well-spread sample (i.e. a sample where the $\pi_{k\ell}$ are small) minimizing the equation \eqref{eq:spatjust}. \citet{graf:lund:13} showed that if the inclusion probabilities are set up proportional to the $\sigma_k$ then \eqref{eq:spatjust} is even more minimized. As result, select a well-spread sample jointly used with the Horvitz-Thompson estimator is a very efficient procedure in terms of variance reduction.


\section{Weakly Associated Vector Sampling}\label{sec:wave}


\subsection{General idea}

Our sampling algorithm, Weakly Associated VEctor (WAVE) sampling starts with the inclusion probability vector. At each step, this vector is randomly modified so that at least one of the components of the vector is replaced by a 0 or a 1. So, in at most $N$ steps a sample is randomly selected. This idea is also used in the cube method proposed by \citet{dev:til:04a} to select balanced samples. The proposed method is different from the cube method by selecting in a completely different way the vector of modification of inclusion probabilities. By carefully choosing the direction of the modification of the working vector, we can ensure that the selection of the sample will be well-spread. This choice is described in Section \ref{sec:implement}.


\subsection{Distance}

In order to describe the spatial structure of the population, a distance is defined as a function $m$ defined on the product set $U\times U$ such that
\begin{equation}\label{eq:dist}
m: U \times U \to \mathbb{R}^{+},
\end{equation}
and satisfies the property of non-negativity, symmetry, and triangular inequality. More specifically, for all $x,y,z \in U$ the following properties hold:
\begin{eqnarray*}
&&m(x,y) \geq 0,~~ m(x,y) = 0 \iff x = y,\\
&&m(x,y) = m(y,x),\\
&&m(x,z) \leq m(x,y) + m(y,z).
\end{eqnarray*}

In most of applications, the usual Euclidean squared distance is used. It is defined by,
\begin{equation}\label{eq:dist_eucli}
m_{E}^2(k,\ell) = (\zb_k - \zb_\ell)^\top(\zb_k - \zb_\ell),
\end{equation}
where $\zb_k$ and $\zb_\ell$ are the spatial coordinates of units $k,\ell\in U$. Sometimes it could be interesting to compute the distance on auxiliary variables. In this case, the Mahalanobis distance can be more appropriate,
$$
m_M^2(k,\ell) = (\xb_k - \xb_\ell)^\top \Sb^{-1} (\xb_k - \xb_\ell),
$$
where
$$
\Sb = \frac{1}{N}\sum_{k\in U} (\xb_k - \bar{\xb})(\xb_k - \bar{\xb})^\top,~~ \bar{\xb} = \frac{1}{N}\sum_{k\in U} \xb_k.
$$

When the population is distributed on a $N_1 \times N_2$ regular grid of $\R^2$, a tore distance can be defined.  We define a tore distance as the Euclidean metric calculated on a regular tore. An advantage of using this distance is that the surface on which we working on, has not anymore corners and edges. With this tore distance, two units on the same column (respectively row) that are on the opposite side have a small distance.  More precisely, a unit that is positioned at the right top corner of the grid will be equally distant to the left top corner and the right bottom corner. It is like seeing the grid curved such that it looks like a regular tore.
 The distance is then defined by:
\begin{equation}\label{eq:dist_tore}
 \begin{array}{lll}
m_T^2(k,\ell) &=& \min[(z_{k1} - z_{\ell 1})^2,(z_{k1} + N_1 - z_{\ell 1})^2,(z_{k1} - N_1 - z_{\ell 1})^2 ]\\
&& + \min[(z_{k2} - z_{\ell 2})^2,(z_{k2} + N_2 - z_{\ell 2})^2,(z_{k2} - N_2 - z_{\ell 2})^2 ]
\end{array} \end{equation}

\begin{exmpl}\label{example:distance}
Let $\{1,\dots,9\}$ be on a regular grid of size $3 \times 3$, then the squared distance matrices defined by Equations~\eqref{eq:dist_eucli} and~\eqref{eq:dist_tore} are equal to
\begin{equation}\label{eq:example_matrices}
 \Mb_{E} = \left(\begin{matrix}
0&    1&    4&    1 &   2 &   5 &   4 &   5 &   8\\
1&    0&    1&    2 &   1 &   2 &   5 &   4 &   5\\
4&    1&    0&    5 &   2 &   1 &   8 &   5 &   4\\
1&    2&    5&    0 &   1 &   4 &   1 &   2 &   5\\
2&    1&    2&    1 &   0 &   1 &   2 &   1 &   2\\
5&    2&    1&    4 &   1 &   0 &   5 &   2 &   1\\
4&    5&    8&    1 &   2 &   5 &   0 &   1 &   4\\
5&    4&    5&    2 &   1 &   2 &   1 &   0 &   1\\
8&    5&    4&    5&    2 &   1 &   4 &   1 &   0\\
\end{matrix}\right), ~~\Mb_{T} = \left(\begin{matrix}
	0&    1 &   1  &  1  &  2  &  2 &   1  &  2  &  2\\
	1&    0 &   1 &   2  &  1  &  2 &   2  &  1  &  2\\
	1 &   1 &   0 &   2  &  2  &  1 &   2  &  2  &  1\\
	1 &   2  &  2 &   0  &  1  &  1 &   1  &  2  &  2\\
	2 &   1 &   2 &   1  &  0  &  1 &   2  &  1  &  2\\
	2 &   2 &   1 &   1  &  1  &  0 &   2  &  2  &  1\\
	1 &   2 &   2 &   1  &  2  &  2 &   0  &  1  &  1\\
	2 &   1 &   2 &   2  &  1  &  2 &   1  &  0  &  1\\
	2 &   2 &   1 &   2  &  2  &  1 &   1  &  1  &  0\\
\end{matrix}\right).\end{equation}
\end{exmpl}

In spatial configuration of a regular grid, some distances between points are equal. The rank of the nearest neighbours is then assigned and duplicated values appear.
In order to obtain a different rank distance for each unit, a small random quantity is added to the coordinates so that it disturbs the given units and the distances are a little bit different from each other.
Let $\varepsilong \in \R^2$ and $\tilde{\zb}_k = \zb _k+ \varepsilong$ the shifted coordinates, Equation~\eqref{eq:dist_tore} is then replaced by,
\begin{equation*}\label{eq:dist_shift}
\begin{array}{lll}
m_S^2(k,\ell) &=& \min[(\tilde{z}_{k1} - z_{\ell 1})^2,(\tilde{z}_{k1} + N_1 - z_{\ell 1})^2,(\tilde{z}_{k1} - N_1 - z_{\ell 1})^2 ]\\
&& + \min[(\tilde{z}_{k2} - z_{\ell 2})^2,(\tilde{z}_{k2} + N_2 - z_{\ell 2})^2,(\tilde{z}_{k2} - N_2 - z_{\ell 2})^2 ].
\end{array}
\end{equation*}
$\varepsilong$ is called a ``shift'' and $m_S$ the shifted version of $m_T$, for example if $\varepsilong = (1/12,1/4)$, the distance matrix $\Mb_{S}$ becomes,
\begin{equation}\label{eq:shift}\Mb_{S}= \left(\begin{matrix}
0 & 0.90 & 1.24 & 0.57 & 1.40 & 1.74 & 1.57 & 2.40 & 2.74 \\
1.24 & 0 & 0.90 & 1.74 & 0.57 & 1.40 & 2.74 & 1.57 & 2.40 \\
0.90 & 1.24 & 0 & 1.40 & 1.74 & 0.57 & 2.40 & 2.74 & 1.57 \\
1.57 & 2.40 & 2.74 & 0 & 0.90 & 1.24 & 0.57 & 1.40 & 1.74 \\
2.74 & 1.57 & 2.40 & 1.24 & 0 & 0.90 & 1.74 & 0.57 & 1.40 \\
2.40 & 2.74 & 1.57 & 0.90 & 1.24 & 0 & 1.40 & 1.74 & 0.57 \\
0.57 & 1.40 & 1.74 & 1.57 & 2.40 & 2.74 & 0 & 0.90 & 1.24 \\
1.74 & 0.57 & 1.40 & 2.74 & 1.57 & 2.40 & 1.24 & 0 & 0.90 \\
1.40 & 1.74 & 0.57 & 2.40 & 2.74 & 1.57 & 0.90 & 1.24 & 0 \\
\end{matrix}\right).
\end{equation}

\begin{figure}[ht!]
 	\centering
 	\input{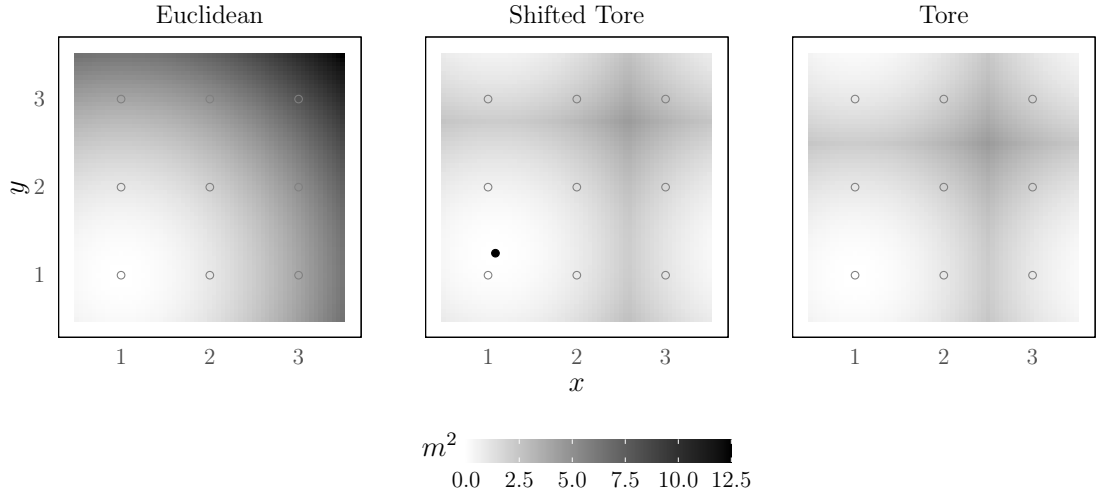}
 	\caption{Simple example of a $3\times 3$ regular grid set up on three different distances with a gradient calculated from the points $(1,1)$.
 The left one is the classical Euclidean distance~\eqref{eq:dist_eucli}, the right one is the tore distance given in~\eqref{eq:dist_tore} and the central graph is the shifted tore distance with a shift equal to $(1/12,1/4)$ (the black point on the graph). It illustrates the two different patterns and the values of the grid points corresponding to the entries of the first row of the three previous matrices~\eqref{eq:example_matrices}.}
 	\label{fig:distance}
\end{figure}

The matrix is no longer a distance matrix since the symmetric axiom has been dropped. A distance that has an unsatisfied symmetry axiom is called a quasi-metric.
Nevertheless, if an epsilon value is added instead of $(1/12,1/4)$, then the values are almost the same and the order is preserved in each row.
In Figure~\ref{fig:distance}, three simple configurations are presented: Euclidean, tore and shifted tore distance on a $3 \times 3$ regular grid.
In shifted distance graph, all the distances from point $(1,1)$ to the other grid points are different.

\subsection{The stratification matrix}\label{sec:stratif}

Let $k\in U$ a unit in the population. The idea is to construct a strata $G_k$ under some distance metric such that the elements in $G_k$ are ranked in increasing order. Define $G_k$ the set of the nearest neighbours of unit $k$, including $k$, such that their inclusion probabilities are greater or equal than one by only one unit. Denote $g_k$ the number of elements inside $G_k$, the spatial weights are then defined as follows

\begin{equation}\label{eq:wpik}
w_{k\ell} =
\left\{
\begin{array}{ll}
 \pi_\ell & \text{ if unit } \ell \text{ is in the set of the } g_k - 1 \text{ nearest neighbour of } k,\\
 \displaystyle\pi_\ell +  1 - \sum_{j\in G_k} \pi_k   & \text{ if unit } \ell \text{ is the } g_k\text{th} \text{ nearest neighbour of } k,\\
 0 & \text{ otherwise.}
 \end{array}
 \right.
 \end{equation}

$\Wb$ denote an $N\times N$ stratification matrix and each row of matrix $\Wb$ represents a stratum. Each stratum is defined by a particular unit and its neighbouring units.
Nearest neighbours are defined with a metric function~\eqref{eq:dist}. If the metric is such that there exists ties values, then we can divide the quantity $w_{kl}$ into the different $g_k$ nearest neighbours of the unit $k$ that have the same distance. Or, a shifted metrics can be used (exemplified in matrix~\eqref{eq:shift}) such that all the distances are different. Each row of matrix $\Wb$ sum to 1. Thus matrix $\Wb$ is a right stochastic matrix. Most of the components of matrix $\Wb$ are null. $\Wb$ can thus be encoded as a sparse matrix.

\begin{exmpl}
Let $U = \{1,2,3,4,5\}$ a population of 5 units. Suppose that the inclusion probabilities are equal to $\pig = (1/2,1/3,1/4,1/5,1/6)$ and that the order in terms of distance metric from the unit 1 is exactly equal to ${1,2,3,4,5}$. Meaning that the 5th unit is the farthest to the first. Then $G_k = \{1,2,3\}$ because $1/2 + 1/3 + 1/4 \cong 1.084 > 1$ and $w_{13} = 1/4 + 1 - (1/2 + 1/3 + 1/4) = 1/6$.
\end{exmpl}

\begin{figure}[ht!]
	\centering
	\input{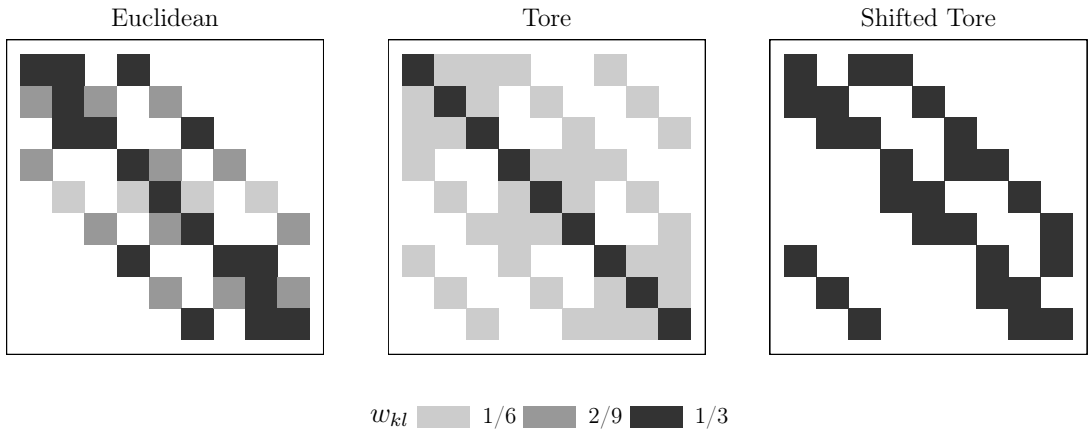}
	\caption{Sparsity pattern of three stratification matrices. Spatial coordinates are $3 \times 3$ regular grid and the inclusion probabilities are equal to $\pi = (1/3,\dots, 1/3)$.
Depending on the way of defining the nearest neighbours in Equation~\eqref{eq:wpik}, different weight values are obtained. The left stratification matrix uses the classical Euclidean distance~\eqref{eq:dist_eucli}, the central one the tore distance~\eqref{eq:dist_tore} and the right one uses a shifted tore distance with a shift randomly generated from a random variable $\mathcal{N}(0,1/100\Ib)$. }
	\label{fig:strat}
\end{figure}

 \begin{exmpl}

Let $\{1,\dots,9\}$ be on a regular grid of size $3 \times 3$ with inclusion probabilities equal $\pi_k=1/3,$ for all $k\in U$.
Figure~\ref{fig:strat} shows different stratification matrices corresponding to $\Mb_{E}$, $\Mb_{T}$ and $\Mb_{S}$ with a shift randomly generated from a random variable $\mathcal{N}(0,1/100\Ib)$ where $\Ib$ is the identity matrix.

\end{exmpl}

Let now $\Db=\diag(\pig)$ the matrix with inclusion probabilities on the diagonal and define $\Ab$ by
\begin{equation}\label{eq:Wsp}
\begin{array}{lll}
\Ab &=& \Wb \Db^{-1}
=
\begin{pmatrix}
w_{11}/\pi_{1} & w_{12}/\pi_{2}  & \cdots & w_{1N}/\pi_{N} \\
\vdots & \vdots & \ddots & \vdots \\
w_{N1}/\pi_{1} & w_{N2}/\pi_{2} & \cdots & w_{NN}/\pi_{N}\\
\end{pmatrix}\\
\end{array}.
\end{equation}

\begin{figure}[ht!]
	\centering
  \input{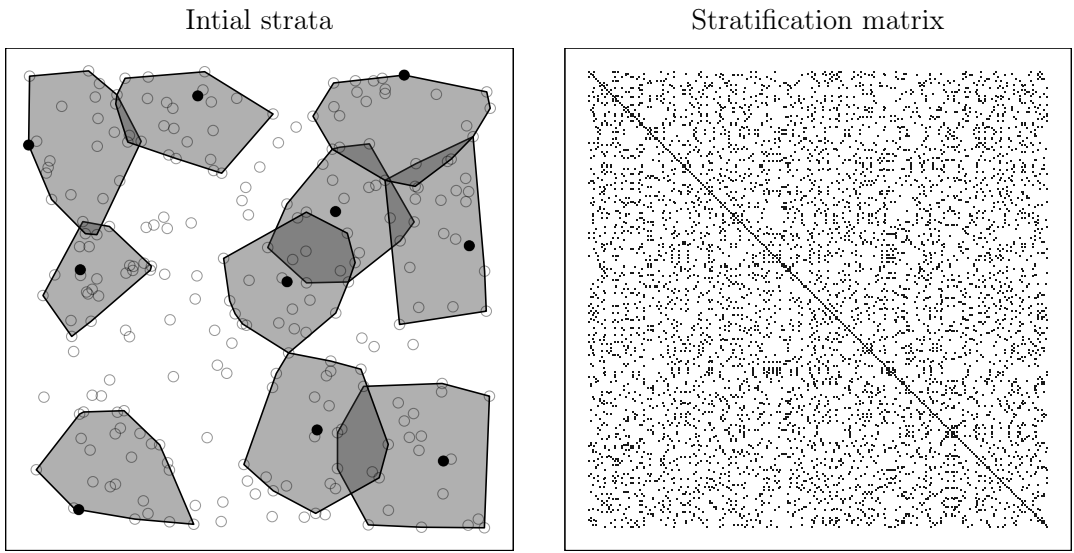}
	\caption{Representation of the strata defined by the spatial weights Equation~\eqref{eq:wpik}. Spatial coordinates of the units are generated randomly from a uniform distribution on the square unit $[0,1] \times [0,1]$. The overall population size is equal to $N = 250$ and the inclusion probabilities are identical and equal to $\pi_k = 1/25 = 0.04$. Meaning that the sample size is equal to $n = 10$. With these parameters the expected number of units in each stratum is equal to $125/4 = 25$. The left graph shows the population and the selected units with its initial strata. On the right, it shows the sparsity pattern of the matrix~\eqref{eq:Wsp}. All entries of the matrix are equal to 1.}
	\label{fig:matrix}
\end{figure}

Matrices $\Wb$ and $\Ab$ are square but not necessarily full rank. The sum of the rows of $\Ab$ is equal or approximately equal to the number of elements in each stratum. The strata are represented by the rows and the contribution of a unit $i$ in each stratum is represented by the $i$th column. Figure~\ref{fig:matrix} shows the sparsity pattern of the two stratification matrices.

\begin{exmpl} Let $U$ be a population of size $N = 250$ and inclusion probabilities equal to $\pi_k=1/25,$ for all $k\in U$. Suppose that spatial coordinates are generated independently from a uniform distribution on the square unit, so that with probability one there are no tied distance values. Since all $1/\pi_k=25$ the non-zero entries of $\Ab$ are all equal to 1. Based on the definition~\eqref{eq:wpik}, the weights are all equal to the inclusion probabilities or zero. Figure~\ref{fig:matrix} shows the sparsity pattern of the stratification matrices and exemplifies some initial strata.
\end{exmpl}


\subsection{Implementation}\label{sec:implement}

The method is described in detail in Algorithm~\ref{algo:1}. The main idea is derived from the cube method \citep{dev:til:04a}. At each step, vector $\pig$ is randomly modified. To modify $\pig$, we choose a vector that spreads at best. Ideally, the aim consists of obtaining a sample $\ab$ such that the following equality is satisfied:
$$
\Ab \ab = \Ab \pig = \textbf{1}.
$$
This linear system define an affine subspace of $\mathbb{R}^N$:
$$
\mathcal{A} = \{\ab \in \mathbb{R}^N \mid \Ab\ab = \Ab\pig \}
$$
which could also be rewrite:
$$
\mathcal{A} = \pig + \text{Null}(\Ab)
$$
where
$$
\text{Null}(\Ab) = \{\vb \in \R^N \mid \Ab\vb = \0b \}.
$$
Depending if matrix $\Ab$ is full rank or not, the vector giving the direction is not selected in the same way.
If matrix $\Ab$ is not full rank, a vector that is contained in the right null space is selected. If matrix $\Ab$ is full rank, we compute $\vb$,$\ub$ a left and a right singular vectors associated to the smallest singular value $\sigma$ of $\Ab$ i.e,
$$
\Ab\vb = \sigma\ub,~~ \Ab^\top\ub = \sigma\vb.
$$
By choosing the modification vector $\vb$, we ensure that we select the vector which remains closest to the set $\mathcal{A}$. Vector $\vb$ is called the weakest associated vector to the matrix $\Ab$. Vector $\vb$ is then centered to ensure the fixed sample size. By using these weakest associated vectors, the initial spatial configurations are the least modified. At each step, some inclusion probabilities $\pig$ are modified and at least one component is set to 0 or 1. Matrix $\Ab$ is updated from the new inclusion probabilities. This step is repeated until there is only one component that is not equal to 0 or 1.

\begin{algorithm}[htb!]
\caption{Algorithm for WAVE sampling}\label{algo:1}
Let  $\Ab = \Ab_0$  and $ \pig_0  = (\pi_1^{(0)}, \dots, \pi_N^{(0)}) = \pig$ for the initialization step. For $t = 0,1,2,\dots$
\begin{enumerate}
	\item From $\pig_t$, extract $\widetilde{\pig}_t$ vector $\pig_t$ restricted to the $k$ such that $ 0 < \pi_{k}^{(t)} <1 $. Let $J$ be the length of $\widetilde{\pig}_t$.

	\item Compute the $J\times J$ matrix $\Ab_t$ of Equation~\eqref{eq:Wsp} using inclusion probabilities $\widetilde{\pig}_t$.

	\item Calculate the rank $r$ of matrix $\Ab_t$.
	\begin{enumerate}
		\item If matrix $\Ab_t$ does not have full rank, choose $\vb_t = (v_1^{(t)},\dots,v_J^{(t)}) \in \R^J$ a vector in the right null space of $\Ab_t$.
		\item If matrix $\Ab_t$ has full rank, compute the singular value decomposition and seek for $\vb_t$ a right singular vector associated to the smallest singular value $\sigma_t$.
	\end{enumerate}
	\item Next in order to ensure the fixed sample size, vector $\vb_t$ is centered:
	$$
    \widetilde{\vb}_t = \displaystyle \vb_t - \frac{1}{J}\sum_{i \in J} v_i^{(t)} \1b_J,
    $$
	where $\1b_J $ is the $J\times 1$ vector of one.
	\item Find $\lambda_1$ and $\lambda_2$ the largest positive real numbers such that all the $ 0\leq \widetilde{\pi}_k^{(t)} + \lambda_1 \widetilde{v}_k^{(t)} \leq 1$ and $ 0\leq \widetilde{\pi}_k^{(t)} - \lambda_2 \widetilde{v}_k^{(t)} \leq 1$, $k = 1,\dots,J$.
	\item Compute
	$$\pig_{t+1} = \left\{\begin{array}{cll}
	\widetilde{\pig}_t + \lambda_1\widetilde{\vb}_t~~ \text{ with probability } \lambda_2/(\lambda_1 + \lambda_2) \\
	\widetilde{\pig}_t - \lambda_2\widetilde{\vb}_t~~ \text{ with probability } \lambda_1/(\lambda_1 + \lambda_2) .\\
	\end{array}\right.$$
	\item Return at 1. with $\pig_{t+1}$ until no units $k$ remains such that $ 0 < \pi_{k}^{(t+1)} <1 $.
\end{enumerate}
\end{algorithm}

Algorithm \ref{algo:1} is implemented in a \texttt{R} package, which uses the Armadillo C++ library into the \texttt{R} interface \citep{eddl:sand:2014}. The implementation uses the sparse matrix class. Indeed, depending on the inclusion probabilities, matrix $\Ab$ given in~\eqref{eq:Wsp} could be strongly sparse. Even if the function benefits from the C++ implementation, it could be quite time consuming as the size of the population $N$ increases. Nevertheless, we will see in the next section that the algorithm performs better in terms of two spreading measures than those currently used for the spatial balanced sampling design.


\section{Spatial balance}\label{sec:index}


\subsection{Vorono{\"\i} polygons}

 \citet{Stev:Olse:spat:2004} suggested the spatial balance of a sample consists of using the Vorono{\"\i} polygons. The Vorono{\"\i} polygon associated to the sample unit $k$ is the set of all units of the population that are closer to $k$ than to any other sample units. Let $v_k$ be the sum of inclusion probabilities of the units belonging to the Vorono{\"\i} polygon associated with the sample unit $k$. If the sample is perfectly spreaded, $v_k$ should be equal to 1 for each $k$.
Indeed,  $n$ units are selected in the sample, then
$$
\sum_{k\in S} v_k = \sum_{k\in U} \pi_k = n,
$$
and so
$$
\frac{1}{n}\sum_{k\in S} v_k = 1.
$$
The variance of the $\textrm{E}[v_k]$ could be approximated and give a good measure of the spatial balance of the sample. The spatial balance measure based on the Vorono{\"\i} polygons is defined by
\begin{equation}\label{eq:voro}
B(S) = \frac{1}{n}\sum_{k\in S}(v_k - 1)^2.
\end{equation}

\begin{figure}[ht!]
\centering
\input{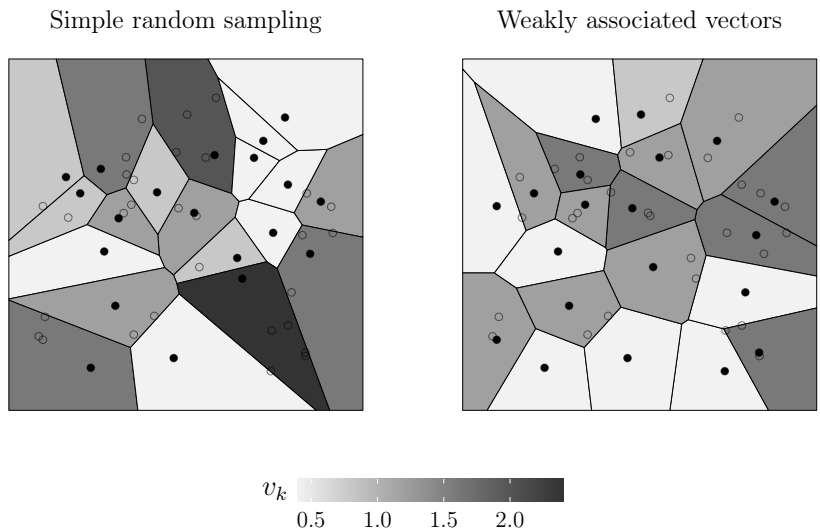}
\caption{Illustrated example of how the spatial balance measure based on the Vorono{\"\i} is performed. The population and sample sizes are respectively equal to $N = 50$ and $n=20$, the inclusion probabilities are identical and equal to $\pi_k = 0.4$. The spatial coordinates are generated from two random uniform $\mathcal{U}(0,1)$. Two sampling design are compared. The left one is the simple random sampling without replacement and the right one is the weakly associated vector sampling.}
\label{fig:voronoi}
\end{figure}

Two samples are compared in Fig.~\ref{fig:voronoi}. The left one is selected with a simple random sampling without replacement and the right one is selected with WAVE sampling. The darker the Vorono{\"\i} polygon, the more units it contains. An exactly well-spread sample should have all polygons of the same colour.

The measure $B$ has some limitations. It does not vary from a fixed finite range. This does not allow a clear understanding if the sample is balanced or clustered \citep{til:dic:esp:giu:18}. Moreover, the measure behaves sometimes wrongly and suggest a well-spread sample although it is not the case. Examples are given in Supplementary Material Section. For these reasons, we suggest to use another measure based on Moran's $I$ index.

\subsection{Moran's $I$ index}\label{moran}

A second approach for measuring the spatial balance of a sampling design has been proposed by \citet{til:dic:esp:giu:18}. Consider a $N\times N$ spatial weights matrix,

$$
\Wb =
\begin{pmatrix}
0      & w_{12} & \cdots & w_{1N} \\
w_{21} & 0      & \cdots & w_{2N} \\
\vdots & \vdots & \ddots & \vdots \\
w_{N1} & w_{N2} & \cdots & 0      \\
\end{pmatrix}.
$$
A large value of $w_{k\ell}$ indicates that $\ell$ is a neighbour of $k$. Matrix $\Wb$ is not necessarily symmetric. The index proposed by \citet{til:dic:esp:giu:18} is defined by
\begin{equation}\label{eq:moran}
I_B(\aaag) = \frac{(\aaag - \bar{\aaag}_w)^\top \Wb (\aaag - \bar{\aaag}_w) }{\sqrt{(\aaag - \bar{\aaag}_w)^\top \Db (\aaag - \bar{\aaag}_w)(\aaag - \bar{\aaag}_w)^\top \Bb (\aaag - \bar{\aaag}_w) }},
\end{equation}
where $\aaag$  is the sample and
$$
\bar{\aaag}_w = \frac{\aaag^\top \Wb \1b}{\1b^\top \Wb \1b},
$$
$\Db$ is the diagonal matrix containing $w_{k.} = \sum_{\ell\in U} w_{k\ell} $ on its diagonal,
$$
\Bb = \Cb^\top\Db\Cb,~~~~ \Cb = \Db^{-1}\Wb - \frac{\1b\1b^\top\Wb}{\1b^\top\Wb\1b},
$$
 and $\1b$ is a column vector of $N$ ones. \citet{til:dic:esp:giu:18} pointed out that $I_B$ can be interpreted as weighted correlation between $\aaa_k$ and the average of the $\aaa_\ell$ that are in the neighbouring of $k$. We have that $-1 \leq I_B \leq 1$ and $I_B = -1$ when the sample is well-spread. \citet{til:dic:esp:giu:18} have proposed to use the inverse of the inclusion probability $h_k = 1/\pi_k$ to define the neighbours of the unit $k$. More specifically, if the unit $k$ is selected it seems natural to consider $h_k -1$ neighbours in the population.
 Let $\lfloor h_k \rfloor$ and $\lceil h_k \rceil$ be respectively the inferior and superior integers of $h_k$. Spatial weights are then defined as follows,
\begin{equation}\label{eq:wpik1}
 w_{k\ell} = \left\{
\begin{array}{ll}
1 & \text{ if  unit } \ell \text{ is in the set of the } \lfloor h_k \rfloor \text{ nearest neighbour of } k\\
h_k - \lfloor h_k \rfloor & \text{ if unit } \ell \text{ is the } \lceil h_k \rceil\text{th} \text{ nearest neighbour of } k \\
0 & \text{ otherwise .}
\end{array}
\right.
 \end{equation}

For example, if a unit $k$ has an inclusion probability of $\pi_k = 0.35$ then $h_k \cong 2.857$. Meaning that the first nearest neighbour of $k$ has a weight equal to 1 and the second has a weight of $0.857$. In case there are units that are at equal distance from each other, \citet{til:dic:esp:giu:18} suggests to divide the spatial weights equally among them.

We propose a new way of defining the spatial weights. It consists of using spatial weights defined in~\eqref{eq:wpik} rather than the weights ~\eqref{eq:wpik1}. We set $w_{kk} = 0$ for all $k \in U$.
For the rest of the paper, $I_B{_1}$ will represent the measure based on the spatial weights~\eqref{eq:wpik1} and $I_B$ the one based on~\eqref{eq:wpik}.


\section{Variance estimation}\label{sec:variance}

If the sampling design is of fixed size, the variance of the Horvitz-Thompson estimator of the total~\eqref{eq:HT} is defined by
\begin{equation*}
\text{var}(\widehat{Y}_{HT})  = -\frac{1}{2}\sum_{k\in U}\sum_{\ell \in U} \left(\frac{y_k}{\pi_k} - \frac{y_\ell}{\pi_\ell}\right)^2\Delta_{k\ell},
\end{equation*}
where $\Delta_{k\ell} = \pi_{k\ell} - \pi_{k}\pi_{\ell}$ and $\pi_{k\ell} = \E(a_ka_l)$ is the joint inclusion probabilities.
For complex sampling designs, quantities $\pi_{k\ell}$ are generally impossible to compute.

Many different estimators have been developed. \citet{sen:53} and \citet{yat:gru:53} proposed one classical estimator:
\begin{equation*}
v_{SYG}(\widehat{Y}_{HT})  =   -\frac{1}{2}\sum_{k\in S}\sum_{\ell \in S} \left(\frac{y_k}{\pi_k} - \frac{y_\ell}{\pi_\ell}\right)^2\frac{\Delta_{k\ell}}{\pi_{k\ell}}.
\end{equation*}
This estimator can take negative values, but it is non-negative when $\Delta_{k\ell} \leq 0$ for all $k \neq \ell \in U$.
A common problem with spatially balanced sampling designs is that many joint inclusion probabilities are equal to zero. Indeed the probability of selecting two close units is generally zero or very close to zero. In this case, $v_{SYG}$ is not an unbiased estimator of $\text{var}(\widehat{Y}_{HT})$.

\citet[][Chapter~5]{till:2020} gives a general estimator based on the variance estimator of the conditional Poisson sampling. It is equal to
\begin{equation*}
v(\widehat{Y}_{HT}) = \sum_{k \in S}\frac{c_k}{\pi_k^2}(y_k - \hat{y}_k^\star)^2,
\end{equation*}
where
$$
\hat{y}_k^\star = \pi_k\frac{\sum_{\ell\in S}c_\ell y_\ell/\pi_\ell}{\sum_{\ell\in S} c_\ell}.
$$
Choosing $c_\ell = (1-\pi_k)n/n-1$ we obtain the H\'ajek-Ros\'en estimator \citet{haj:81} defined by
\begin{equation}\label{eq:varHAJ}
v_{HAJ}(\widehat{Y}_{HT}) = \frac{n}{n-1}\sum_{k\in S}(1-\pi_k)\left\{\frac{y_k}{\pi_k} - \frac{\sum_{\ell\in S} y_\ell(1-\pi_\ell)/\pi_\ell}{\sum_{\ell\in S} (1-\pi_\ell)} \right\}^2.
\end{equation}

This variance estimator is simple to compute and has the advantage of using only the first-order inclusion probabilities. It is a good estimator for maximum entropy sampling design and simple random sampling without replacement. \citet{gra:lun:sch:12} pointed out that the estimator seems to overestimate the variance for spread sampling design. \citet{gra:sch:14} proposed an estimator based on the nearest neighbour in the sample. It is called variance estimator for spatially balanced sampled and is defined as follow:
\begin{equation}\label{eq:varSB}
v_{SB}(\widehat{Y}_{HT}) = \frac{1}{2}\sum_{k\in S}\left(\frac{y_k}{\pi_k} - \frac{y_{\ell_k}}{\pi_{\ell_k}}\right)^2,
\end{equation}
where $\ell_k$ is the nearest neighbour to the unit $k$ in the sample. \citet{Stev:Olse:vari:2003} proposed an estimator based on a local neighbourhood for each units in the sample. It is called the local mean variance estimator and is given by
\begin{equation}\label{eq:varLM}
v_{LM}(\widehat{Y}_{HT}) = \sum_{k\in U}\sum_{\ell \in D_k} w_{k\ell}\left( \frac{y_k}{\pi_k} - \sum_{m\in D_k} w_{km}\frac{y_m}{\pi_m} \right)^2,
\end{equation}
where the weights $w_{k\ell}$ are computed such that they vary inversely as $\pi_\ell$ and decrease as the distance between unit $k$ and $\ell$ increases. Moreover, it satisfies the constraint $\sum_{k \in S} w_{k\ell} = \sum_{\ell \in S} w_{k\ell} = 1$. The set $D_k$ is the neighbourhood of the unit $k$ and is defined by the unit itself and the three neighbourhoods of the three nearest neighbours. Meaning that $D_k$ contains at least four units and at most thirteen.
This variance estimator is implemented by function \texttt{localmean.var} in the R package ``spsurvey'' \citet{spsurvey}. It produces a good estimator for the GRTS method. For the rest of the manuscript, we will adopt the following notation: $v_{LM_j}(\widehat{Y}_{HT})$ where $j$ is the number of neighbours used in the calculations. In Section \ref{sec:realdata}, we compare the previous estimators for different sampling designs.


\section{Simulations on artificial spatial configurations}\label{sec:artificial}

In this section, we propose three artificial spatial configurations to study the performance of the WAVE sampling in terms of spreading measure. To generate the three population datasets, the expected size of the population is equal to $N=144$.

\begin{enumerate}
	\item The dataset is generated from the Complete Spatial Randomness (CSR) that is a Poisson process with intensity equal to $N$, meaning that the expected number of points in the unit square is equal to $N$.
	\item A Neyman-Scott cluster process \citep{ney:scot:1958} is generated with 12 circular discs of radius 0.055 with units uniformly distributed around the centre. Each cluster contains 12 units such that the population target size is equal to $N$.
	\item  Simple regular grid of size $12\times 12$.
\end{enumerate}

\begin{figure}[ht!]
	\centering
	\input{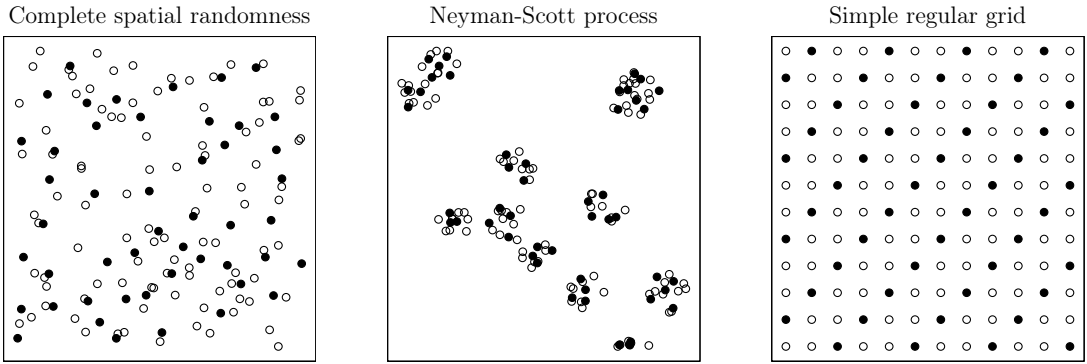}
	\caption{Example of a sample selection by the WAVE sampling on the three different spatial configurations, Complete Spatial Randomness, Neyman-Scott and regular grid. For each of them, the inclusion probabilities are equal to $\pi_k = n/N$ for all $k\in U$.}
	\label{fig:artificialExample2}
\end{figure}

 Figure~\ref{fig:artificialExample2} shows a sample selection by the WAVE sampling design on the three different datasets. For the three configurations, the sample size is equal to $n = $ 3 and the inclusion probabilities are all equal to $\pi_k = n/N$ for all $k\in U$. When units are regularly dispersed in the space and when the inverse of inclusion probabilities is equal to an integer that is a divisor of the population size $N$, the selected sample can be systematic, which is the optimal solution.

For each population, 10,000 samples of size $n$ respectively equal to 25, 50 and 100 are selected. Two cases are considered for the inclusion probabilities. In the first case, all inclusion probabilities are equal
$$
\pi_k = \pi = \frac{n}{N}, \text{ for all } k \in U.
$$
For the second case, the inclusion probabilities are unequal and sum up to $n$,
$$
\text{ we have } \pi_k \neq \pi_\ell \text{ and } \sum_{k \in U} \pi_k = n, \text{ for all } k,\ell \in U, k \neq \ell
$$

In each case we calculate the spatial balance based on the Vorono{\"\i} polygons~\eqref{eq:voro} and measures based on Moran's $I$ index~\eqref{eq:moran}. The simulation results of the CRS dataset are given in the Table~\ref{tab:tablepp}. For the measures based on Moran's $I$ index, the WAVE sampling design performs better than the other algorithms. Moreover, for the classical measure based on the Vorono{\"\i} polygons, the WAVE sampling design performs equally and sometimes better than the local pivotal method.
This can be explained by the fact that the spatial balance measure based on the Vorono{\"\i} polygons is less sensitive to observe a well-spread sample and sometimes suggest a well-spread sample although it is not the case (See Supplementary Material Section). For the equal probabilities designs the measures $I_{B_1}$ and $I_B$ coincide.
Indeed the strata based on the inverse inclusion probabilities are the same as the ones considered such that the inclusion probabilities sum to 1.
For unequal sampling designs, the differences are less marked with the measure based on the inverse inclusion probabilities~\eqref{eq:wpik1}. This result comes from the heterogeneity of the strata and the randomness of the algorithm. If the inclusion probabilities of a unit is nearly zero, then the size of the strata will be very large. This effect can increase the spatial balance measure. Similar results for the two remaining datasets can be seen in the Supplementary Material Section. This analysis shows that the measure $I_B$ should be prefered to $I_{B_1}$.

\begin{knitrout}
\definecolor{shadecolor}{rgb}{0.969, 0.969, 0.969}\color{fgcolor}\begin{table}

\caption{\label{tab:tablepp}Spreading measures results based on 10000 simulations on the Complete spatial randomness dataset. The population size is equal to 144.}
\centering
\resizebox{\linewidth}{!}{
\fontsize{9}{11}\selectfont
\begin{tabular}[t]{lrrrrrrrrrrr}
\toprule
\multicolumn{1}{c}{ } & \multicolumn{11}{c}{Sampling design} \\
\cmidrule(l{3pt}r{3pt}){2-12}
\multicolumn{1}{c}{ } & \multicolumn{6}{c}{Equal probabilities} & \multicolumn{5}{c}{Unequal probabilities} \\
\cmidrule(l{3pt}r{3pt}){2-7} \cmidrule(l{3pt}r{3pt}){8-12}
  & wave & lpm1 & scps & grts & hip & srswor & wave & lpm1 & scps & grts & maxent\\
\midrule
\addlinespace[1ex]
\multicolumn{12}{l}{$I_{B_1}$}\\
\hspace{1em}$n = 16 $ & -0.530 & -0.348 & -0.370 & -0.220 & -0.259 & -0.030 & -0.028 & -0.009 & -0.012 & 0.027 & 0.093\\
\hspace{1em}$n = 32 $ & -0.693 & -0.467 & -0.464 & -0.322 & -0.392 & -0.017 & -0.125 & -0.095 & -0.085 & -0.059 & 0.016\\
\hspace{1em}$n = 48 $ & -0.807 & -0.583 & -0.506 & -0.375 & -0.373 & -0.015 & -0.436 & -0.344 & -0.318 & -0.229 & -0.020\\
\addlinespace[1ex]
\multicolumn{12}{l}{$I_B$}\\
\hspace{1em}$n = 16 $ & -0.530 & -0.348 & -0.370 & -0.220 & -0.259 & -0.030 & -0.459 & -0.316 & -0.331 & -0.201 & -0.028\\
\hspace{1em}$n = 32 $ & -0.693 & -0.467 & -0.464 & -0.322 & -0.392 & -0.017 & -0.548 & -0.393 & -0.373 & -0.261 & -0.013\\
\hspace{1em}$n = 48 $ & -0.807 & -0.583 & -0.506 & -0.375 & -0.373 & -0.015 & -0.621 & -0.469 & -0.424 & -0.292 & -0.029\\
\addlinespace[1ex]
\multicolumn{12}{l}{$B$}\\
\hspace{1em}$n = 16 $ & 0.115 & 0.117 & 0.108 & 0.164 & 0.135 & 0.338 & 0.123 & 0.124 & 0.118 & 0.177 & 0.345\\
\hspace{1em}$n = 32 $ & 0.137 & 0.128 & 0.130 & 0.167 & 0.165 & 0.345 & 0.140 & 0.146 & 0.138 & 0.180 & 0.352\\
\hspace{1em}$n = 48 $ & 0.158 & 0.137 & 0.149 & 0.177 & 0.195 & 0.337 & 0.165 & 0.151 & 0.158 & 0.189 & 0.319\\
\bottomrule
\end{tabular}}
\end{table}

\end{knitrout}


\section{Application to the Meuse dataset}\label{sec:realdata}

This section investigates the application of WAVE sampling on the dataset ``Meuse'' available in the R package ``sp'' of \citet{pebe:biva:2005}. It is described as follows:
``This data set gives locations and topsoil heavy metal concentrations, along with a number of soil and landscape variables at the observation locations, collected in a flood plain of the river Meuse, near the village of Stein (NL). Heavy metal concentrations are from composite samples of an area of approximately 15 m x 15 m.''

In order to see how the WAVE sampling performs in terms of spread measures, $m = 10,000$ samples of size respectively equal to 15, 30 and 50 are selected. As in the previous simulation with an artificial population, two cases are considered, equal and unequal probabilities. In the latter case, inclusion probabilities are set proportional to concentration of copper. Locations with high concentrations of copper were therefore more likely to be selected into the sample. Let $Y$ be the total cadmium concentration over the whole population.
To show that the variance of the estimated total with the WAVE sampling design is lower than the other method, we calculate the approximated variance with the following quantity:
\begin{equation}\label{eq:varSIM}
v_{SIM}(\widehat{Y}_{HT}) = \frac{1}{m}\sum_{s} \left\{\widehat{Y}_{HT}(s) - Y\right\}^2.
\end{equation}

\begin{figure}[ht]
	\centering
	\input{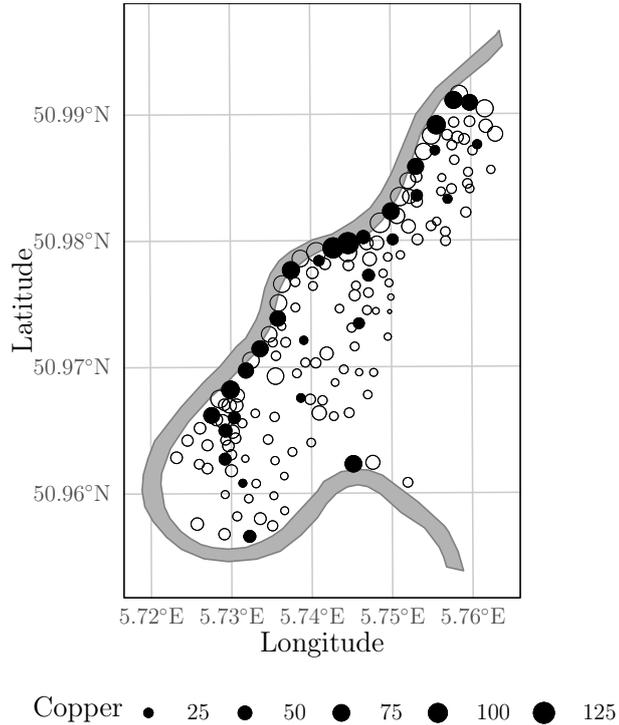}
	\caption{Example of WAVE sampling on the Meuse dataset. The overall population size is equal to 155. The inclusion probabilities are proportional to copper level variable and the sample size is equal to 30. Plotted sizes of the units are proportional to the copper concentration.
The Meuse River is filled in light blue.}
	\label{fig:swiss}
\end{figure}

Figure~\ref{fig:swiss} shows sample selected with the WAVE sampling. The filled black circles are selected units while the hollow circles are those that are not selected in the sample.
We observe that the dataset is partially aggregated around the river showing a strong spatial correlation.

Results of the three spatial balanced measures on 10'000 simulated samples is given in Table~\ref{tab:spread_meuse_table}. WAVE sampling performs better than other sampling designs in terms of $I_B$ and $I_{B_1}$.
In terms of spatial balance measure $B$, the algorithms are comparable to the artificial simulation, the differences are less marked.

Results of the simulations on the variance estimator in Table~\ref{tab:V_ht_meuse} shows that the WAVE sampling strategy has a lower variance than the currently used method.
This suggests that the method is more efficient in cases where there is a clear spatial correlation. A design-unbiased variance estimator does not exist for the Horvitz-Thompson estimator, but the spatially balanced estimator ~\eqref{eq:varSB} seems to produce a good estimator for this dataset. Although the latter slightly overestimates the variance none of the other estimators seem to offer a better alternative. As there is no unbiased estimator we favour a slight overestimation of the variance. Table~\ref{tab:tablevar} shows the coverage rate as well as the ratio $v_{SB}/v
_{SIM}$ for all sampling methods.

Based on these simulation results, we are confident that we propose here a new method that allows to select a sample with a really strong degree of spreading. It performs better than the other sampling method. It can be generalized to higher dimensions and respects the unequal inclusion probabilities.

\begin{table}

\caption{\label{tab:spread_meuse_table}Spreading measures results based on 10000 simulations on the Meuse dataset. The population size is equal to 155.}
\centering
\resizebox{\linewidth}{!}{
\fontsize{9}{11}\selectfont
\begin{tabular}[t]{lrrrrrrrrrrr}
\toprule
\multicolumn{1}{c}{ } & \multicolumn{11}{c}{Sampling design} \\
\cmidrule(l{3pt}r{3pt}){2-12}
\multicolumn{1}{c}{ } & \multicolumn{6}{c}{Equal probabilities} & \multicolumn{5}{c}{Unequal probabilities} \\
\cmidrule(l{3pt}r{3pt}){2-7} \cmidrule(l{3pt}r{3pt}){8-12}
  & wave & lpm1 & scps & grts & hip & srswor & wave & lpm1 & scps & grts & maxent\\
\midrule
\addlinespace[1ex]
\multicolumn{12}{l}{$I_{B_1}$}\\
\hspace{1em}$n = 15 $ & -0.518 & -0.338 & -0.351 & -0.226 & -0.230 & -0.030 & -0.340 & -0.250 & -0.246 & -0.165 & -0.003\\
\hspace{1em}$n = 30 $ & -0.664 & -0.427 & -0.427 & -0.266 & -0.259 & -0.019 & -0.407 & -0.298 & -0.288 & -0.172 & 0.024\\
\hspace{1em}$n = 50 $ & -0.796 & -0.519 & -0.473 & -0.302 & -0.248 & -0.011 & -0.466 & -0.326 & -0.285 & -0.204 & 0.038\\
\addlinespace[1ex]
\multicolumn{12}{l}{$I_B$}\\
\hspace{1em}$n = 15 $ & -0.518 & -0.338 & -0.351 & -0.226 & -0.230 & -0.030 & -0.354 & -0.244 & -0.247 & -0.153 & 0.009\\
\hspace{1em}$n = 30 $ & -0.664 & -0.427 & -0.427 & -0.266 & -0.259 & -0.019 & -0.427 & -0.290 & -0.283 & -0.154 & 0.048\\
\hspace{1em}$n = 50 $ & -0.796 & -0.519 & -0.473 & -0.302 & -0.248 & -0.011 & -0.455 & -0.305 & -0.263 & -0.181 & 0.060\\
\addlinespace[1ex]
\multicolumn{12}{l}{$B$}\\
\hspace{1em}$n = 15 $ & 0.119 & 0.125 & 0.118 & 0.170 & 0.160 & 0.379 & 0.115 & 0.121 & 0.120 & 0.170 & 0.387\\
\hspace{1em}$n = 30 $ & 0.118 & 0.123 & 0.126 & 0.164 & 0.159 & 0.359 & 0.120 & 0.121 & 0.120 & 0.162 & 0.345\\
\hspace{1em}$n = 50 $ & 0.139 & 0.132 & 0.143 & 0.174 & 0.194 & 0.329 & 0.138 & 0.133 & 0.141 & 0.160 & 0.281\\
\bottomrule
\end{tabular}}
\end{table}

\begin{knitrout}
\definecolor{shadecolor}{rgb}{0.969, 0.969, 0.969}\color{fgcolor}\begin{table}

\caption{\label{tab:tablevar}Results of 10000 simulations on Meuse dataset. The population size is equal to 155. $v_{SIM}$ is equal to the variance approximated by the simulations \eqref{eq:varSIM}. $v$ depends on the sampling design.
For the srswor and maxent methods, we used the estimator $v_{HAJ}$ \eqref{eq:varHAJ} while for the other sampling designs, we use $v_{SB}$ \eqref{eq:varSB}. Coverage rate of the 95\% confidence intervals are computed as well as the ratio between averages of $v$ and $v_{SIM}$.}
\centering
\resizebox{\linewidth}{!}{
\fontsize{9}{11}\selectfont
\begin{tabular}[t]{lrrrrrrrrrrr}
\toprule
\multicolumn{1}{c}{ } & \multicolumn{11}{c}{Sampling design} \\
\cmidrule(l{3pt}r{3pt}){2-12}
\multicolumn{1}{c}{ } & \multicolumn{6}{c}{Equal probabilities} & \multicolumn{5}{c}{Unequal probabilities} \\
\cmidrule(l{3pt}r{3pt}){2-7} \cmidrule(l{3pt}r{3pt}){8-12}
  & wave & lpm1 & scps & grts & hip & srswor & wave & lpm1 & scps & grts & maxent\\
\midrule
\addlinespace[1em]
\multicolumn{12}{l}{$v_{SIM}$}\\
\hspace{1em}$n = 15 $ & 1.232 & 1.387 & 1.309 & 1.517 & 1.315 & 1.774 & 0.250 & 0.287 & 0.260 & 0.330 & 0.361\\
\hspace{1em}$n = 30 $ & 0.533 & 0.525 & 0.538 & 0.586 & 0.463 & 0.805 & 0.116 & 0.109 & 0.096 & 0.115 & 0.150\\
\hspace{1em}$n = 50 $ & 0.250 & 0.250 & 0.222 & 0.284 & 0.200 & 0.413 & 0.052 & 0.049 & 0.039 & 0.049 & 0.065\\
\addlinespace[1em]
\multicolumn{12}{l}{$v$}\\
\hspace{1em}$n = 15 $ & 1.847 & 1.670 & 1.635 & 1.596 & 1.701 & 1.784 & 0.393 & 0.362 & 0.371 & 0.333 & 0.365\\
\hspace{1em}$n = 30 $ & 0.692 & 0.687 & 0.670 & 0.657 & 0.639 & 0.808 & 0.154 & 0.153 & 0.152 & 0.150 & 0.153\\
\hspace{1em}$n = 50 $ & 0.380 & 0.375 & 0.385 & 0.353 & 0.337 & 0.403 & 0.081 & 0.078 & 0.080 & 0.080 & 0.066\\
\addlinespace[1em]
\multicolumn{12}{l}{Coverage of the 95\% confidence interval}\\
\hspace{1em}$n = 15 $ & 0.925 & 0.907 & 0.914 & 0.887 & 0.918 & 0.890 & 0.973 & 0.958 & 0.972 & 0.929 & 0.933\\
\hspace{1em}$n = 30 $ & 0.953 & 0.943 & 0.942 & 0.929 & 0.963 & 0.924 & 0.971 & 0.972 & 0.983 & 0.966 & 0.942\\
\hspace{1em}$n = 50 $ & 0.975 & 0.966 & 0.977 & 0.946 & 0.973 & 0.927 & 0.978 & 0.979 & 0.990 & 0.979 & 0.944\\
\addlinespace[1em]
\multicolumn{12}{l}{Ratio $v/v_{SIM}$}\\
\hspace{1em}$n = 15 $ & 1.499 & 1.204 & 1.249 & 1.052 & 1.294 & 1.006 & 1.573 & 1.264 & 1.428 & 1.011 & 1.011\\
\hspace{1em}$n = 30 $ & 1.298 & 1.307 & 1.246 & 1.121 & 1.380 & 1.003 & 1.323 & 1.400 & 1.588 & 1.308 & 1.016\\
\hspace{1em}$n = 50 $ & 1.521 & 1.501 & 1.739 & 1.242 & 1.685 & 0.976 & 1.564 & 1.615 & 2.030 & 1.616 & 1.003\\
\bottomrule
\end{tabular}}
\end{table}

\end{knitrout}

\begin{table}

\caption{\label{tab:V_ht_meuse}Results of 10000 simulations on Meuse dataset. The population size is equal to 155. $v_{SIM}$ \eqref{eq:varSIM} is equal to the variance approximated by the simulations. $v_{SB}$ \eqref{eq:varSB} is the variance estimator based on the nearest neighbours in the sample. $v_{LM_j}$ is equal to the estimator \eqref{eq:varLM} where the number of neighbouring units used is set to $j = 2,3,4$. $v_{HAJ}$ \eqref{eq:varHAJ} is the Hajek-Rosen estimator.}
\centering
\resizebox{\linewidth}{!}{
\fontsize{9}{11}\selectfont
\begin{tabular}[t]{lrrrrrrrrrrr}
\toprule
\multicolumn{1}{c}{ } & \multicolumn{11}{c}{Sampling design} \\
\cmidrule(l{3pt}r{3pt}){2-12}
\multicolumn{1}{c}{ } & \multicolumn{6}{c}{Equal probabilities} & \multicolumn{5}{c}{Unequal probabilities} \\
\cmidrule(l{3pt}r{3pt}){2-7} \cmidrule(l{3pt}r{3pt}){8-12}
  & wave & lpm1 & scps & grts & hip & srswor & wave & lpm1 & scps & grts & maxent\\
\midrule
\addlinespace[1ex]
\multicolumn{12}{l}{$n = 15$}\\
\hspace{1em}$v_{SIM}$ & 1.232 & 1.387 & 1.309 & 1.517 & 1.315 & 1.774 & 0.250 & 0.287 & 0.260 & 0.330 & 0.361\\
\hspace{1em}$v_{SB}$ & 1.847 & 1.670 & 1.635 & 1.596 & 1.701 & 1.455 & 0.393 & 0.362 & 0.371 & 0.333 & 0.321\\
\hspace{1em}$v_{LM2}$ & 0.962 & 0.889 & 0.889 & 0.855 & 0.930 & 0.786 & 0.224 & 0.206 & 0.209 & 0.194 & 0.183\\
\hspace{1em}$v_{LM3}$ & 1.301 & 1.256 & 1.261 & 1.230 & 1.308 & 1.147 & 0.293 & 0.279 & 0.282 & 0.269 & 0.259\\
\hspace{1em}$v_{LM4}$ & 1.463 & 1.445 & 1.452 & 1.430 & 1.487 & 1.352 & 0.325 & 0.315 & 0.319 & 0.306 & 0.299\\
\hspace{1em}$v_{HAJ}$ & 1.808 & 1.824 & 1.829 & 1.826 & 1.854 & 1.784 & 0.375 & 0.370 & 0.373 & 0.369 & 0.365\\
\addlinespace[1ex]
\multicolumn{12}{l}{$n = 30$}\\
\hspace{1em}$v_{SIM}$ & 0.533 & 0.525 & 0.538 & 0.586 & 0.463 & 0.805 & 0.116 & 0.109 & 0.096 & 0.115 & 0.150\\
\hspace{1em}$v_{SB}$ & 0.692 & 0.687 & 0.670 & 0.657 & 0.639 & 0.634 & 0.154 & 0.153 & 0.152 & 0.150 & 0.143\\
\hspace{1em}$v_{LM2}$ & 0.382 & 0.373 & 0.370 & 0.362 & 0.356 & 0.348 & 0.094 & 0.090 & 0.090 & 0.089 & 0.082\\
\hspace{1em}$v_{LM3}$ & 0.555 & 0.543 & 0.543 & 0.534 & 0.534 & 0.512 & 0.130 & 0.127 & 0.127 & 0.126 & 0.118\\
\hspace{1em}$v_{LM4}$ & 0.654 & 0.649 & 0.649 & 0.641 & 0.652 & 0.616 & 0.150 & 0.148 & 0.148 & 0.147 & 0.140\\
\hspace{1em}$v_{HAJ}$ & 0.808 & 0.805 & 0.806 & 0.808 & 0.814 & 0.808 & 0.153 & 0.154 & 0.155 & 0.154 & 0.153\\
\addlinespace[1ex]
\multicolumn{12}{l}{$n = 50$}\\
\hspace{1em}$v_{SIM}$ & 0.250 & 0.250 & 0.222 & 0.284 & 0.200 & 0.413 & 0.052 & 0.049 & 0.039 & 0.049 & 0.065\\
\hspace{1em}$v_{SB}$ & 0.380 & 0.375 & 0.385 & 0.353 & 0.337 & 0.344 & 0.081 & 0.078 & 0.080 & 0.080 & 0.080\\
\hspace{1em}$v_{LM2}$ & 0.214 & 0.208 & 0.213 & 0.196 & 0.190 & 0.190 & 0.050 & 0.048 & 0.049 & 0.048 & 0.045\\
\hspace{1em}$v_{LM3}$ & 0.308 & 0.294 & 0.298 & 0.284 & 0.280 & 0.276 & 0.068 & 0.068 & 0.069 & 0.068 & 0.065\\
\hspace{1em}$v_{LM4}$ & 0.358 & 0.349 & 0.351 & 0.340 & 0.337 & 0.330 & 0.079 & 0.080 & 0.081 & 0.081 & 0.078\\
\hspace{1em}$v_{HAJ}$ & 0.406 & 0.407 & 0.407 & 0.404 & 0.405 & 0.403 & 0.065 & 0.066 & 0.066 & 0.066 & 0.066\\
\bottomrule
\end{tabular}}
\end{table}

\clearpage

\section{Discussion}

Environmental data are generally not uniformly distributed over a region of the space. Thus it is generally justified to use unequal inclusion probabilities to overrepresent some parts of the population. As explained in Section \ref{sec:spatjust}, this reduces the variance of the Horvitz-Thompson estimator, a phenomenon also observed in Section \ref{sec:realdata} on the Meuse dataset.

In this manuscript, we present a sampling design that selects the units in a very well-spread configuration. We have shown on the Meuse dataset that on measurements of spatial spreading the method behaves very well. Moreover, the approximated variance of the Horvitz-Thompson estimator is lower with WAVE sampling than the other methods. Some second-order inclusion probabilities are null. It is thus impossible to estimate unbiasedly the variance of the estimator. However, we propose different estimators and compare their performance. We show that it is possible to estimate appropriately the variance and to construct confidence intervals that have good coverage rates, particularly when the sample size is large. All of these results indicate that our method is very efficient to select a well-spread sample and has better properties than the usual spatial sampling designs.


\newpage
\section*{Acknowledgments}

We would like to thank the associate editor and two reviewers for their conscientious reading and positive comments, which improved the quality of this manuscript. We are grateful to Pierre-Yves Deléamont, Ziqing Dong, Esther Eustache, Cliona Jauslin and Lionel Qualité for their time spent to bring valuable comments at an early stage of this manuscript.


\bibliographystyle{apalike}

\end{document}